\documentclass[11pt]{article}

\usepackage[margin=2.5cm]{geometry}
\usepackage{amsmath,amssymb}
\usepackage{graphicx}
\usepackage{authblk}
\usepackage[colorlinks=true,linkcolor=blue,citecolor=blue,urlcolor=blue]{hyperref}

\newcommand{\Vt}{\acute{V}}
\newcommand{\xit}{\acute{\xi}}

\title{Exact analytical spectra of an effective field-modified radial Schr\"odinger equation by the Nikiforov--Uvarov method}

\author[1]{Tunde Joseph Osunmusanmi\thanks{Corresponding author. Email: \href{mailto:Tunde.josunmusanmi@ku.ac.ae}{Tunde.josunmusanmi@ku.ac.ae}}}
\author[1,2]{Berihu Teklu}
\affil[1]{College of Computing and Mathematical Sciences, Department of Applied Mathematics and Sciences, Khalifa University of Science and Technology, 127788 Abu Dhabi, United Arab Emirates}
\affil[2]{ KU Research Center for Intelligent Computing, Networks and Sensing (ICONS), Khalifa University of Science and Technology, 127788,
Abu Dhabi, United Arab Emirates}
\affil[ ]{\textit{Email:} \href{mailto:berihu.gebrehiwot@ku.ac.ae}{berihu.gebrehiwot@ku.ac.ae} (B. Teklu)}
\date{}

\begin{document}

\maketitle

\begin{abstract}
We revisit the effective radial Schr\"odinger equation for a charged spinless particle in collinear static electric and magnetic fields that was formulated in our earlier work within the tridiagonal representation approach (TRA). In that model the angular factor of the electric coupling is replaced by the orientation parameter $C=\cos\theta$, the linear Zeeman term is diagonal in the azimuthal quantum number, and the effective field-modified central potential takes the form $V(r)=\Vt(r)+qECr$. Here we show that the same equation can be solved exactly without the TRA, by the Nikiforov--Uvarov (NU) method, once the potential component $\Vt(r)$ is prescribed. Three examples are worked out in closed form: a Kratzer-type component, which yields a linear-plus-Coulomb-plus-inverse-square central potential; a pseudoharmonic-type component, which yields a linear-plus-harmonic-plus-inverse-square potential; and a Hulth\'en-type component with a repulsive core (Manning--Rosen form), which yields a screened confining potential with a finite number of $s$-wave bound states. For each case we obtain the energy spectrum, the normalized eigenfunctions in terms of Laguerre or Jacobi polynomials, the location of the potential minimum, and closed-form radial moments. The spectra are independent of the orientation $\theta$ and of the electric field, in agreement with the structure found by the TRA, while the field enters through the shape of $V(r)$, the radial moments, and the expectation value $\langle V\rangle_k$. All analytical results are confirmed by an independent finite-difference diagonalization.


\end{abstract}

%=====================================================================
\section{Introduction}
\label{sec:intro}
%=====================================================================

The response of bound quantum systems to static external fields is a classic problem of atomic, molecular and mesoscopic physics. Electric fields mix states of different angular momentum, magnetic fields produce Zeeman shifts and, at larger strengths, diamagnetic confinement, and the full three-dimensional problem in combined fields is generally nonseparable~\cite{Bethe1957,Landau1977}. Reduced radial models that retain the dominant field-induced structure while preserving a self-adjoint Sturm--Liouville form are therefore useful as analytical benchmarks~\cite{Paris2014}.

In a previous paper~\cite{Osunmusanmi2026} we formulated such a reduced description: an effective radial Schr\"odinger equation for a charged spinless particle in collinear uniform electric and magnetic fields, obtained under a weak-magnetic-field condition and an angular projection in which the factor $\cos\theta$ of the electric coupling is replaced by an orientation parameter $C=\cos\theta$. We emphasized there that the resulting model is not the exact Stark--Zeeman problem but a field-aligned effective radial model. The equation was solved in Ref.~\cite{Osunmusanmi2026} by the tridiagonal representation approach (TRA)~\cite{Alhaidari2017,Alhaidari2019,Alhaidari2020}, which in its search mode determines the potentials that admit a tridiagonal matrix representation in a chosen basis. In the Laguerre basis this led to linear-plus-inverse and linear-plus-quadratic field-modified central potentials, and in the Jacobi basis to a singular confining potential treated by a generalized eigenvalue problem. The TRA, together with the closely related J-matrix method~\cite{Heller1974,YamaniFishman1975,JmatrixBook2008,Ismail2011}, has been used extensively to construct exactly and quasi-exactly solvable systems and to engineer potential functions with prescribed spectra~\cite{AlhaidariBahlouli2008,AlhaidariIsmail2015,AlhaidariTaiwo2017,Taiwo2018,Taiwo2019,Taiwo2020,Alhaidari2021,Alhaidari2023,Alhaidari2025}. In this program the expansion coefficients of the wavefunction are orthogonal polynomials in the energy and physical parameters, several of which are new to the mathematical literature~\cite{Alhaidari2020,Alhaidari2018}.

A natural question is whether the effective equation of Ref.~\cite{Osunmusanmi2026} is tied to the TRA, or whether it can be treated with other exact analytical techniques. In this paper we answer the second part affirmatively. We do not use the TRA here. Instead, we prescribe the potential component $\Vt(r)$ of the effective central potential and solve the resulting equation with the Nikiforov--Uvarov (NU) method~\cite{Nikiforov1988,Tezcan2009,Berkdemir2006}, which reduces a second-order equation of hypergeometric type to a polynomial eigenvalue problem through a Rodrigues formula. The NU method belongs to the broad family of exact techniques for radial problems, which ranges from the factorization method~\cite{InfeldHull1951} and the classification of potentials solvable in terms of hypergeometric functions~\cite{Natanzon1979,Ginocchio1984} to shape-invariant supersymmetric quantum mechanics~\cite{Cooper1995,Levai1989}; see also Ref.~\cite{Dong2011} for higher-dimensional extensions. The purpose is to show that the effective model admits exact solutions by standard analytical methods, and to enlarge the family of field-modified central potentials with closed-form spectra and eigenfunctions.

We present three examples. The first uses a Kratzer-type component~\cite{Kratzer1920,Bayrak2007} and produces a central potential combining linear, Coulomb and inverse-square terms, which is related in form to the Cornell potential~\cite{Eichten1978}. The second uses a pseudoharmonic-type component~\cite{Davidson1932,Sage1985,Ikhdair2007} and produces a linear-plus-harmonic-plus-inverse-square potential. The third uses a Hulth\'en-type component with a repulsive core~\cite{Hulthen1942,Flugge1974,ManningRosen1933} and produces a screened confining potential with a finite $s$-wave spectrum. In all three cases the spectra are independent of the orientation angle and of the electric field, exactly as in the TRA construction, while the field enters through the shape of the central potential, the position of its minimum, and the radial moments of the eigenstates, all of which we obtain analytically.

We state the contribution of this work explicitly. The energy spectra obtained below are those of the prescribed components $\Vt(r)$; for the Kratzer, pseudoharmonic and Manning--Rosen forms these spectra are known in the literature~\cite{Bayrak2007,Ikhdair2007,ManningRosen1933}, and we do not claim them as new. The new elements are the following: (i) the orientation-dependent families of effective central potentials $V(r)=\Vt(r)+qEC\,r$ generated from these components, together with an analytical characterization of their geometry, including the existence, location and depth of the potential minimum and their dependence on $\theta$; (ii) closed-form radial moments of the eigenstates, obtained from the Laguerre and Jacobi algebra and the Hellmann--Feynman theorem, and the orientation-dependent expectation values constructed from them; and (iii) the demonstration that the effective model of Ref.~\cite{Osunmusanmi2026} is not tied to the TRA but admits exact treatment by an independent analytical method.

The paper is organized as follows. Section~\ref{sec:model} recalls the effective radial equation of Ref.~\cite{Osunmusanmi2026}. Section~\ref{sec:NU} summarizes the NU method. Sections~\ref{sec:kratzer}, \ref{sec:pseudo} and \ref{sec:hulthen} contain the three examples. Section~\ref{sec:results} presents the numerical validation and discusses the results, and Sec.~\ref{sec:conclusion} concludes. A worked example of the NU method, for the Kratzer-type component, is given in the Appendix.

%=====================================================================
\section{The effective radial model}
\label{sec:model}
%=====================================================================

We briefly recall the construction of Ref.~\cite{Osunmusanmi2026}, to which we refer for details and for the validity conditions. A nonrelativistic spinless particle of mass $m$ and charge $q$ moves in a central potential and in uniform fields $\mathbf{B}=B\hat{\mathbf{z}}$ and $\mathbf{E}=E\hat{\mathbf{z}}$. In the symmetric gauge, and with the diamagnetic term neglected under the weak-magnetic-field condition of Ref.~\cite{Osunmusanmi2026}, the Hamiltonian contains the linear Zeeman term $-(qB/2m)L_z$ and the electric dipole coupling, which is linear in $r\cos\theta$. The angular factor is then replaced by the orientation parameter
\begin{equation}
C=\cos\theta,\qquad 0\le\theta\le\frac{\pi}{2},
\label{eq:C}
\end{equation}
and the azimuthal dependence $e^{im_\phi\phi}$ diagonalizes the linear Zeeman term. In atomic units, $\hbar=m=1$, writing $\Psi=\allowbreak r^{-1}\psi(r)Y_{lm_\phi}(\theta,\phi)$, the reduced radial equation reads~\cite{Osunmusanmi2026}
\begin{equation}
\left[-\frac{1}{2}\frac{d^{2}}{dr^{2}}+\frac{l(l+1)}{2r^{2}}+\Vt(r)\right]\psi(r)=\xit\,\psi(r),
\label{eq:radial}
\end{equation}
with the boundary conditions $\psi(0)=0$ and $\psi(r)\to 0$ as $r\to\infty$. The effective field-modified central potential is
\begin{equation}
V(r)=\Vt(r)+qEC\,r ,
\label{eq:V}
\end{equation}
where we follow the sign convention of Ref.~\cite{Osunmusanmi2026}, in which $qEC>0$ corresponds to a confining linear term. The eigenvalue $\xit$ is related to the energy $\xi$ by the linear magnetic shift
\begin{equation}
\xit=\xi+\frac{qB}{2}\,m_\phi .
\label{eq:zeeman}
\end{equation}
Consequently, the level shift $\Delta\xi=\xit-\xi=qBm_\phi/2$ is proportional to the magnetic field and independent of $l$ and of the radial quantum number, as found in Ref.~\cite{Osunmusanmi2026}.

In the TRA treatment of Ref.~\cite{Osunmusanmi2026}, $\Vt(r)$ was determined by the requirement of tridiagonality. Here we adopt the converse viewpoint: $\Vt(r)$ is prescribed, Eq.~(\ref{eq:radial}) is solved exactly by the NU method, and Eq.~(\ref{eq:V}) then defines a one-parameter family of field-modified central potentials labeled by $\theta$. Because Eq.~(\ref{eq:radial}) involves only $\Vt(r)$, its spectrum $\xit$ does not depend on $\theta$ or on $E$; this is the same structural property obtained with the TRA. The orientation and the electric field instead control the geometry of $V(r)$ and the expectation values of field-dependent quantities, which we compute below.

It is useful to state precisely what the construction~(\ref{eq:V}) represents. The eigenvalue problem~(\ref{eq:radial}) is posed for $\Vt(r)$, so its eigenvalues and eigenfunctions are fixed by $\Vt(r)$, and the magnetic field adds the Zeeman shift~(\ref{eq:zeeman}). The effective central potential $V(r)$ is the potential obtained by combining this component with the projected linear field term; it is the object whose shape depends on the field and on the orientation. In this formulation the field therefore acts on the potential landscape associated with a given spectrum, rather than on the energy levels themselves, and the quantities that carry the field and orientation dependence are the geometry of $V(r)$ and the mean value of the field term, $qE\cos\theta\,\langle r\rangle_k$, in each eigenstate. A complementary formulation, in which the linear term is retained in the radial operator so that the spectrum itself depends on $qE\cos\theta$, leads to quasi-exactly solvable problems and is treated separately~\cite{Osunmusanmi2026c}.

%=====================================================================
\section{The Nikiforov--Uvarov method}
\label{sec:NU}
%=====================================================================

The NU method~\cite{Nikiforov1988} applies to equations that can be brought, by a change of variable $s=s(r)$, to the generalized hypergeometric form
\begin{equation}
\psi''(s)+\frac{\tilde{\tau}(s)}{\sigma(s)}\psi'(s)+\frac{\tilde{\sigma}(s)}{\sigma^{2}(s)}\psi(s)=0,
\label{eq:NUform}
\end{equation}
where $\sigma$ and $\tilde{\sigma}$ are polynomials of degree at most two and $\tilde{\tau}$ is a polynomial of degree at most one. The factorization $\psi(s)=\varphi(s)y(s)$ with $\varphi'/\varphi=\pi/\sigma$ reduces Eq.~(\ref{eq:NUform}) to the hypergeometric-type equation
\begin{equation}
\sigma(s)y''+\tau(s)y'+\lambda y=0,\qquad \tau=\tilde{\tau}+2\pi ,
\label{eq:hyp}
\end{equation}
where
\begin{equation}
\pi(s)=\frac{\sigma'-\tilde{\tau}}{2}\pm\sqrt{\left(\frac{\sigma'-\tilde{\tau}}{2}\right)^{2}-\tilde{\sigma}+\kappa\sigma},
\qquad \lambda=\kappa+\pi' .
\label{eq:pi}
\end{equation}
The constant $\kappa$ is fixed by requiring the radicand to be a perfect square, so that $\pi(s)$ is a polynomial of degree at most one. Polynomial solutions $y_n(s)$ exist when
\begin{equation}
\lambda=\lambda_n=-n\tau'-\frac{n(n-1)}{2}\sigma'',\qquad n=0,1,2,\dots,
\label{eq:lambdan}
\end{equation}
and they are given by the Rodrigues formula~\cite{Szego1975,Ismail2005,Abramowitz1964}
\begin{equation}
y_n(s)=\frac{B_n}{\rho(s)}\frac{d^{n}}{ds^{n}}\Big[\sigma^{n}(s)\rho(s)\Big],\qquad
\big(\sigma\rho\big)'=\tau\rho .
\label{eq:rodrigues}
\end{equation}
Among the admissible choices of $\pi$, the bound states correspond to the one for which $\tau'<0$, which guarantees normalizability~\cite{Nikiforov1988,Tezcan2009}. Equating $\lambda$ with $\lambda_n$ gives the quantization condition. Other exact approaches, such as supersymmetric quantum mechanics~\cite{Cooper1995} or the asymptotic iteration method~\cite{Ciftci2003}, lead to the same spectra for the potentials considered here; we use the NU method because it delivers the spectrum, the eigenfunctions and the weight function in a single construction.

%=====================================================================
\section{Kratzer-type component}
\label{sec:kratzer}
%=====================================================================

As the first example we take
\begin{equation}
\Vt(r)=-\frac{A}{r}+\frac{G}{r^{2}},\qquad A>0,
\label{eq:VtK}
\end{equation}
which is the Kratzer form~\cite{Berkdemir2006,Kratzer1920}. The corresponding effective field-modified central potential is
\begin{equation}
V(r)=qEC\,r-\frac{A}{r}+\frac{G}{r^{2}} .
\label{eq:VK}
\end{equation}
For $G=0$ it has the same linear-plus-inverse functional form as the first family of Ref.~\cite{Osunmusanmi2026}, and it is related in form to the Cornell potential~\cite{Eichten1978}. The inverse-square term combines with the centrifugal term through the effective angular parameter
\begin{align}
L(L+1)&=l(l+1)+2G,\nonumber\\
L&=-\frac{1}{2}+\sqrt{\left(l+\frac{1}{2}\right)^{2}+2G},
\label{eq:L}
\end{align}
where we require $(l+\tfrac12)^2+2G>0$ and choose the root $L>-\tfrac12$, which gives the regular behavior $\psi\sim r^{L+1}$ at the origin.

\subsection{Spectrum and eigenfunctions}

With $\epsilon^{2}=-2\xit$, Eq.~(\ref{eq:radial}) takes the form~(\ref{eq:NUform}) in the variable $s=r$ with
\begin{equation}
\sigma=r,\qquad \tilde{\tau}=0,\qquad \tilde{\sigma}=-\epsilon^{2}r^{2}+2Ar-L(L+1).
\label{eq:NUK}
\end{equation}
From Eq.~(\ref{eq:pi}),
\begin{equation}
\pi=\frac{1}{2}\pm\sqrt{\epsilon^{2}r^{2}+(\kappa-2A)r+\left(L+\tfrac{1}{2}\right)^{2}} ,
\end{equation}
and the radicand is a perfect square for $\kappa=2A\pm\epsilon(2L+1)$. The choice yielding $\tau'<0$ (see Appendix~\ref{app:kratzer}) is
\begin{gather}
\pi=L+1-\epsilon r,\qquad \tau=2(L+1)-2\epsilon r,\nonumber\\
\kappa=2A-\epsilon(2L+1).
\label{eq:choiceK}
\end{gather}
Then $\lambda=2A-2\epsilon(L+1)$, while Eq.~(\ref{eq:lambdan}) gives $\lambda_k=2\epsilon k$. Equating the two yields $\epsilon_k=A/(k+L+1)$ and therefore
\begin{equation}
\xit_{k,l}=-\frac{A^{2}}{2\left(k+L+1\right)^{2}},\qquad k=0,1,2,\dots
\label{eq:EK}
\end{equation}
The physical energies follow from Eq.~(\ref{eq:zeeman}),
\begin{equation}
\xi_{k,l,m_\phi}=-\frac{A^{2}}{2\left(k+L+1\right)^{2}}-\frac{qB}{2}\,m_\phi .
\label{eq:EKphys}
\end{equation}
The spectrum is infinite, accumulates at zero, and depends on $l$ only through $L$; for $G=0$ it reduces to the hydrogen-like form with principal quantum number $k+l+1$.

The factor $\varphi=r^{L+1}e^{-\epsilon_k r}$ and the weight function $\rho=r^{2L+1}e^{-2\epsilon_k r}$ follow from $\varphi'/\varphi=\pi/\sigma$ and Eq.~(\ref{eq:rodrigues}), and the Rodrigues formula yields associated Laguerre polynomials. The normalized eigenfunctions are
\begin{align}
\psi_{k,l}(r)&=N_{k}\,r^{L+1}e^{-\epsilon_k r}\,L_{k}^{(2L+1)}(2\epsilon_k r),\label{eq:psiK}\\
N_{k}^{2}&=\frac{(2\epsilon_k)^{2L+3}\,k!}{2(k+L+1)\,\Gamma(k+2L+2)} .\label{eq:NK}
\end{align}

\subsection{Field-dependent geometry and radial moments}

Although the spectrum~(\ref{eq:EK}) is independent of $\theta$, the central potential~(\ref{eq:VK}) is not. For $qEC>0$ and $G>0$, the stationarity condition $V'(r)=0$ becomes
\begin{equation}
qEC\,r^{3}+A\,r-2G=0,
\label{eq:rminK}
\end{equation}
which by Descartes' rule of signs has exactly one positive root $r_{\min}(\theta)$. To first order in the field,
\begin{equation}
r_{\min}\simeq\frac{2G}{A}-\frac{8qEC\,G^{3}}{A^{4}},\quad
V_{\min}\simeq-\frac{A^{2}}{4G}+\frac{2qEC\,G}{A},
\label{eq:rminKapprox}
\end{equation}
so that increasing the field projection (decreasing $\theta$) moves the minimum inward and raises it.

Closed-form radial moments follow from the Laguerre algebra and from the Hellmann--Feynman theorem~\cite{Feynman1939}, applied to $A$ and to $G$ (using $dL/dG=1/(L+\tfrac12)$):
\begin{align}
\langle r\rangle_{k}&=\frac{3(k+L+1)^{2}-L(L+1)}{2A},\label{eq:rK}\\
\left\langle r^{-1}\right\rangle_{k}&=\frac{A}{(k+L+1)^{2}},\label{eq:rinvK}\\
\left\langle r^{-2}\right\rangle_{k}&=\frac{A^{2}}{(k+L+1)^{3}\left(L+\tfrac{1}{2}\right)} .\label{eq:rinv2K}
\end{align}
These give the expectation value of the effective field-modified central potential in the $k$th eigenstate,
\begin{align}
\langle V\rangle_{k}={}&-\frac{A^{2}}{(k+L+1)^{2}}+\frac{GA^{2}}{(k+L+1)^{3}\left(L+\tfrac{1}{2}\right)}\nonumber\\
&+qE\cos\theta\,\frac{3(k+L+1)^{2}-L(L+1)}{2A},
\label{eq:VexpK}
\end{align}
whose orientation-dependent part, $qE\cos\theta\,\langle r\rangle_k$, is the mean value of the projected field term in the $k$th eigenstate. We use $\langle V\rangle_k$ as a diagnostic of how the orientation modifies the potential sampled by each state, not as an independent observable. Since $\langle r\rangle_k$ grows as $(k+L+1)^2$, the field contribution dominates for excited states; see Fig.~\ref{fig:Vexp}.

%=====================================================================
\section{Pseudoharmonic-type component}
\label{sec:pseudo}
%=====================================================================

As the second example we take
\begin{equation}
\Vt(r)=\frac{1}{2}\omega^{2}r^{2}+\frac{G}{r^{2}}+c_{0},\qquad \omega>0,
\label{eq:VtP}
\end{equation}
which contains the pseudoharmonic potential $D_0(r/r_0-r_0/r)^2$ as the special case $\omega^2=2D_0/r_0^2$, $G=D_0r_0^2$, $c_0=-2D_0$~\cite{Sage1985,Ikhdair2007}, and the Davidson potential as another~\cite{Davidson1932}. The effective field-modified central potential is
\begin{equation}
V(r)=qEC\,r+\frac{1}{2}\omega^{2}r^{2}+\frac{G}{r^{2}}+c_{0},
\label{eq:VP}
\end{equation}
which for $G=c_0=0$ has the same linear-plus-quadratic form as the second family of Ref.~\cite{Osunmusanmi2026}.

\subsection{Spectrum and eigenfunctions}

With $\varepsilon=\xit-c_0$ and the variable $s=r^{2}$, Eq.~(\ref{eq:radial}) becomes
\begin{equation}
\psi''(s)+\frac{1}{2s}\psi'(s)+\frac{2\varepsilon s-\omega^{2}s^{2}-L(L+1)}{4s^{2}}\psi(s)=0,
\end{equation}
with $L$ given by Eq.~(\ref{eq:L}). This is of the form~(\ref{eq:NUform}) with
\begin{equation}
\sigma=2s,\qquad \tilde{\tau}=1,\qquad \tilde{\sigma}=2\varepsilon s-\omega^{2}s^{2}-L(L+1) .
\end{equation}
Then
\begin{equation}
\pi=\frac{1}{2}\pm\sqrt{\omega^{2}s^{2}+2(\kappa-\varepsilon)s+\left(L+\tfrac{1}{2}\right)^{2}},
\end{equation}
and the choice with $\tau'<0$ is
\begin{gather}
\pi=L+1-\omega s,\qquad \tau=2L+3-2\omega s,\nonumber\\
\kappa=\varepsilon-\omega\left(L+\tfrac{1}{2}\right).
\end{gather}
Hence $\lambda=\varepsilon-\omega(L+\tfrac32)$ and $\lambda_k=2\omega k$, which gives
\begin{equation}
\xit_{k,l}=c_{0}+\omega\left(2k+L+\frac{3}{2}\right),\qquad k=0,1,2,\dots
\label{eq:EP}
\end{equation}
and, with Eq.~(\ref{eq:zeeman}),
\begin{equation}
\xi_{k,l,m_\phi}=c_{0}+\omega\left(2k+L+\frac{3}{2}\right)-\frac{qB}{2}\,m_\phi .
\label{eq:EPphys}
\end{equation}
The spectrum is equally spaced in $k$ with spacing $2\omega$, and $G$ shifts it through $L$. The weight function is $\rho=s^{L+1/2}e^{-\omega s}$, and the normalized eigenfunctions are
\begin{align}
\psi_{k,l}(r)&=N_{k}\,r^{L+1}e^{-\omega r^{2}/2}\,L_{k}^{(L+1/2)}(\omega r^{2}),\label{eq:psiP}\\
N_{k}^{2}&=\frac{2\,\omega^{L+3/2}\,k!}{\Gamma\!\left(k+L+\tfrac{3}{2}\right)} .\label{eq:NP}
\end{align}

\subsection{Field-dependent geometry and radial moments}

For $G>0$ the minimum of $V(r)$ satisfies
\begin{equation}
\omega^{2}r^{4}+qEC\,r^{3}-2G=0,
\label{eq:rminP}
\end{equation}
which again has a unique positive root. Expanding about the field-free value $r_0=(2G/\omega^{2})^{1/4}$,
\begin{equation}
r_{\min}\simeq r_{0}-\frac{qEC}{4\omega^{2}},\qquad
V_{\min}\simeq\omega\sqrt{2G}+c_{0}+qEC\,r_{0}.
\end{equation}
The Hellmann--Feynman theorem with respect to $\omega$ and $G$ gives
\begin{equation}
\langle r^{2}\rangle_{k}=\frac{2k+L+\tfrac{3}{2}}{\omega},\qquad
\langle r^{-2}\rangle_{k}=\frac{\omega}{L+\tfrac{1}{2}},
\label{eq:momP}
\end{equation}
and for the ground state the first moment is
\begin{equation}
\langle r\rangle_{0}=\frac{\Gamma(L+2)}{\Gamma\!\left(L+\tfrac{3}{2}\right)\sqrt{\omega}} .
\label{eq:r0P}
\end{equation}
Therefore
\begin{align}
\langle V\rangle_{0}={}&c_{0}+\frac{\omega}{2}\left(L+\frac{3}{2}\right)+\frac{G\,\omega}{L+\tfrac{1}{2}}\nonumber\\
&+qE\cos\theta\,\frac{\Gamma(L+2)}{\Gamma\!\left(L+\tfrac{3}{2}\right)\sqrt{\omega}} .
\label{eq:VexpP}
\end{align}

%=====================================================================
\section{Hulth\'en-type component with a repulsive core}
\label{sec:hulthen}
%=====================================================================

As the third example we take a Hulth\'en attraction~\cite{Hulthen1942,Flugge1974} supplemented by a short-range repulsive core,
\begin{align}
\Vt(r)={}&-V_{0}\,\frac{e^{-\delta r}}{1-e^{-\delta r}}+V_{1}\,\frac{e^{-2\delta r}}{\left(1-e^{-\delta r}\right)^{2}},\nonumber\\
&V_0,V_1\ge0,\quad \delta>0,
\label{eq:VtH}
\end{align}
which has the Manning--Rosen form~\cite{ManningRosen1933} and reduces to the pure Hulth\'en potential for $V_1=0$. The effective field-modified central potential is
\begin{equation}
V(r)=qEC\,r-V_{0}\,\frac{e^{-\delta r}}{1-e^{-\delta r}}+V_{1}\,\frac{e^{-2\delta r}}{\left(1-e^{-\delta r}\right)^{2}} .
\label{eq:VH}
\end{equation}
Using $e^{-x}/(1-e^{-x})=1/x-\tfrac12+x/12+O(x^{3})$, its short-distance behavior is
\begin{equation}
V(r)\simeq\frac{V_{1}}{\delta^{2}r^{2}}-\frac{V_{0}+V_{1}}{\delta r}+\frac{V_{0}}{2}+\frac{5V_{1}}{12}+O(r),
\label{eq:VHsmall}
\end{equation}
while $V(r)\simeq qEC\,r$ at large distances. Thus $V(r)$ has a repulsive inverse-square core, an intermediate Coulomb-like attraction, and a linear outer wall whose slope is controlled by the orientation $\theta$. We restrict this example to $s$ waves, $l=0$, for which the solution is exact; for $l\neq0$ the centrifugal term is not of the form~(\ref{eq:VtH}) and an approximation of Pekeris~\cite{Pekeris1934} or Greene--Aldrich~\cite{GreeneAldrich1976} type would be required, which we deliberately avoid.

\subsection{Potential minimum}

With $z=e^{-\delta r}\in(0,1)$, the condition $V'(r)=0$ becomes the cubic equation
\begin{equation}
qEC\,(1-z)^{3}+V_{0}\delta\, z(1-z)-2V_{1}\delta\, z^{2}=0 .
\label{eq:rminH}
\end{equation}
Its left-hand side equals $qEC\ge0$ at $z=0$ ($r\to\infty$) and $-2V_1\delta<0$ at $z=1$ ($r\to0$), so for $V_1>0$ a minimum exists for every orientation. In the field-free projection, $\theta=\pi/2$, Eq.~(\ref{eq:rminH}) gives the closed form
\begin{align}
r_{\min}\big|_{\theta=\pi/2}&=\frac{1}{\delta}\ln\!\left(1+\frac{2V_{1}}{V_{0}}\right),\nonumber\\
V_{\min}\big|_{\theta=\pi/2}&=-\frac{V_{0}^{2}}{4V_{1}},
\label{eq:rminH0}
\end{align}
and increasing $qE\cos\theta$ moves the minimum inward and raises it. For $V_1=0$ the core disappears, Eq.~(\ref{eq:rminH}) has no root in $(0,1)$, and $V(r)$ increases monotonically from $-\infty$, as for the Cornell potential.

\subsection{Spectrum and eigenfunctions}

We define
\begin{equation}
\epsilon^{2}=-\frac{2\xit}{\delta^{2}},\quad g=\frac{2V_{0}}{\delta^{2}},\quad h=\frac{2V_{1}}{\delta^{2}},\quad
\eta=\frac{1}{2}+\sqrt{\frac{1}{4}+h},
\label{eq:defsH}
\end{equation}
so that $\eta(\eta-1)=h$ and $\eta\ge1$. In the variable $z$ the $s$-wave equation takes the form~(\ref{eq:NUform}) with
\begin{gather}
\sigma=z(1-z),\qquad \tilde{\tau}=1-z,\nonumber\\
\tilde{\sigma}=-\epsilon^{2}+(2\epsilon^{2}+g)z-(\epsilon^{2}+g+h)z^{2}.
\end{gather}
The radicand in Eq.~(\ref{eq:pi}) is a perfect square for $\kappa=g\pm\epsilon(2\eta-1)$. The choice with $\tau'<0$ is
\begin{gather}
\pi=\epsilon-(\epsilon+\eta)z,\qquad \kappa=g-\epsilon(2\eta-1),\nonumber\\
\tau=1+2\epsilon-(2\epsilon+2\eta+1)z ,
\end{gather}
so that $\lambda=g-\eta-2\epsilon\eta$. Since $\sigma''=-2$, Eq.~(\ref{eq:lambdan}) gives $\lambda_n=n(n+2\epsilon+2\eta)$, and the quantization condition reads $g+h-(n+\eta)^{2}=2\epsilon(n+\eta)$. Hence
\begin{align}
\epsilon_{n}&=\frac{g+h-(n+\eta)^{2}}{2(n+\eta)},\nonumber\\
\xit_{n}&=-\frac{\delta^{2}}{8}\left[\frac{g+h-(n+\eta)^{2}}{n+\eta}\right]^{2},
\label{eq:EH}
\end{align}
and $\xi_{n,m_\phi}=\xit_n-qBm_\phi/2$. Normalizability requires $\epsilon_n>0$, so the bound states are labeled by
\begin{equation}
n=0,1,\dots,n_{\max},\qquad n_{\max}+\eta<\sqrt{g+h},
\label{eq:nmax}
\end{equation}
and the spectrum is finite. For $V_1=0$ one has $\eta=1$, and Eq.~(\ref{eq:EH}) reduces to the Hulth\'en result $\xit_n=-(\delta^{2}/8)\big[(g-(n+1)^{2})/(n+1)\big]^{2}$; in the further limit $\delta\to0$ with $V_0/\delta=A$ fixed it tends to $-A^{2}/[2(n+1)^{2}]$, the $s$-wave limit of Eq.~(\ref{eq:EK}) with $G=0$.

The factor $\varphi=z^{\epsilon_n}(1-z)^{\eta}$ and the weight $\rho=z^{2\epsilon_n}(1-z)^{2\eta-1}$ yield Jacobi polynomials through the Rodrigues formula, and the eigenfunctions are
\begin{align}
\psi_{n}(r)={}&N_{n}\,e^{-\epsilon_{n}\delta r}\left(1-e^{-\delta r}\right)^{\eta}\nonumber\\
&\times P_{n}^{(2\epsilon_{n},\,2\eta-1)}\!\left(1-2e^{-\delta r}\right).
\label{eq:psiH}
\end{align}
Near the origin $\psi_n\propto r^{\eta}$, so the repulsive core acts as an effective centrifugal barrier, with $\eta$ playing the role of $L+1$ in Eq.~(\ref{eq:L}).

\subsection{Ground-state moments}

For $n=0$, with $a=2\epsilon_0$, the substitution $r=-\delta^{-1}\ln z$ reduces the normalization and first-moment integrals to Euler beta integrals and their parameter derivatives:
\begin{align}
N_{0}^{2}&=\frac{\delta}{\mathrm{B}(a,\,2\eta+1)},\label{eq:N0H}\\
\langle r\rangle_{0}&=\frac{1}{\delta}\Big[\psi_{\Gamma}(a+2\eta+1)-\psi_{\Gamma}(a)\Big],\label{eq:r0H}
\end{align}
where $\mathrm{B}$ is the Euler beta function and $\psi_{\Gamma}$ the digamma function. For $V_1=0$ ($\eta=1$), Eq.~(\ref{eq:r0H}) becomes $\langle r\rangle_0=\delta^{-1}\big[a^{-1}+(a+1)^{-1}+(a+2)^{-1}\big]$. The field-dependent part of the expectation value of the central potential in the ground state is $qE\cos\theta\,\langle r\rangle_0$.

%=====================================================================
\section{Results and discussion}
\label{sec:results}
%=====================================================================

As a consistency check of the analytical formulas, we diagonalized Eq.~(\ref{eq:radial}) independently by a second-order finite-difference discretization on a large uniform grid. The parameters are $A=1$, $G=1/2$ for the Kratzer-type case; $\omega=1$, $G=1/2$, $c_0=0$ for the pseudoharmonic-type case; and $V_0=1/2$, $V_1=1/20$, $\delta=1/4$ (so that $g=16$, $h=8/5$, $\eta\simeq1.860$) for the Hulth\'en-type case. Table~\ref{tab:levels} compares the closed-form eigenvalues with the numerical ones; the agreement is at the level of the discretization error in all cases. For the Hulth\'en-type case Eq.~(\ref{eq:nmax}) gives $n+\eta<\sqrt{g+h}\simeq4.195$, i.e., exactly three bound states, $n=0,1,2$, which is what the numerical diagonalization finds. The first moments~(\ref{eq:rK}), (\ref{eq:r0P}) and (\ref{eq:r0H}), and the root of Eq.~(\ref{eq:rminH}), were checked in the same way.

\begin{table}[t]
\caption{Closed-form eigenvalues $\xit$ of Eq.~(\ref{eq:radial}) compared with finite-difference (FD) diagonalization. Kratzer-type: $A=1$, $G=1/2$; pseudoharmonic-type: $\omega=1$, $G=1/2$, $c_0=0$; Hulth\'en-type ($l=0$): $V_0=1/2$, $V_1=1/20$, $\delta=1/4$. The values are independent of $\theta$ and $E$.}
\label{tab:levels}
\centering
\begin{tabular*}{0.75\textwidth}{@{\extracolsep\fill}cccc@{\extracolsep\fill}}
\hline
Case & $(l,k)$ & Analytical & FD \\
\hline
Kratzer & $(0,0)$ & $-0.19098301$ & $-0.19098373$ \\
        & $(0,1)$ & $-0.07294902$ & $-0.07294923$ \\
        & $(0,2)$ & $-0.03819660$ & $-0.03819669$ \\
        & $(1,0)$ & $-0.09429024$ & $-0.09429026$ \\
        & $(1,1)$ & $-0.04583654$ & $-0.04583656$ \\
\hline
Pseudoharmonic & $(0,0)$ & $\phantom{-}2.11803399$ & $\phantom{-}2.11803398$ \\
               & $(0,1)$ & $\phantom{-}4.11803399$ & $\phantom{-}4.11803394$ \\
               & $(1,0)$ & $\phantom{-}2.80277564$ & $\phantom{-}2.80277563$ \\
               & $(1,1)$ & $\phantom{-}4.80277564$ & $\phantom{-}4.80277561$ \\
\hline
Hulth\'en & $(0,0)$ & $-0.45142464$ & $-0.45142511$ \\
          & $(0,1)$ & $-0.08473726$ & $-0.08473741$ \\
          & $(0,2)$ & $-0.00382009$ & $-0.00382011$ \\
\hline
\end{tabular*}

\end{table}

\begin{figure}[t]
\centering
\includegraphics[width=\textwidth]{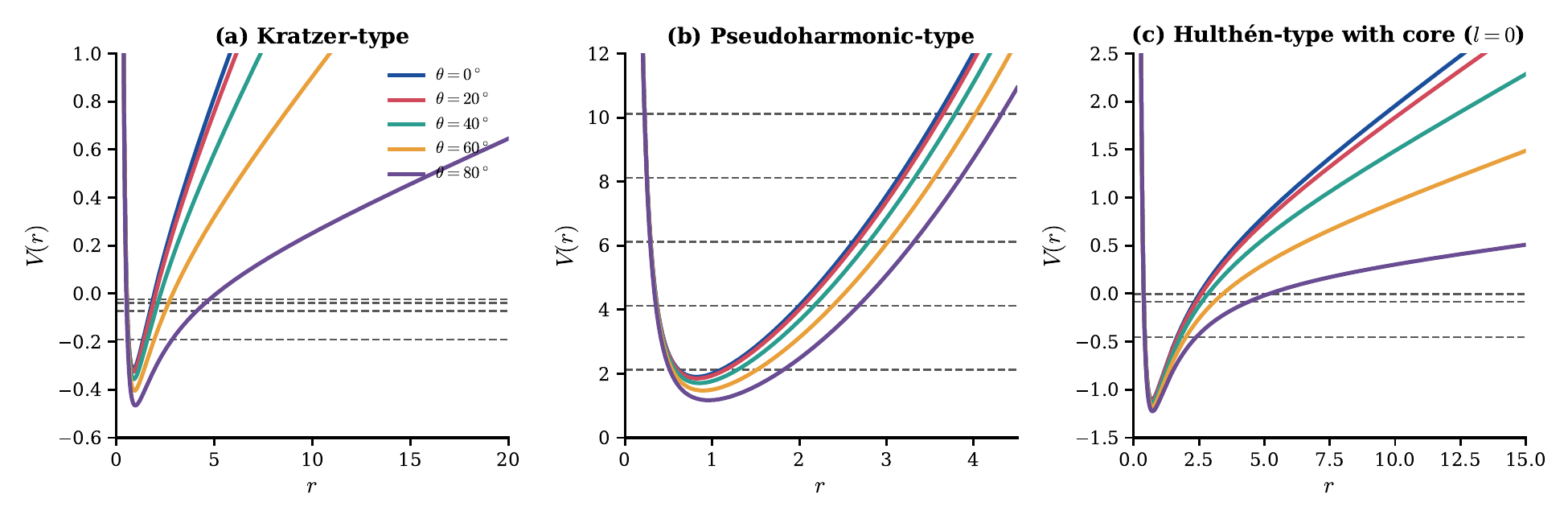}
\caption{Effective field-modified central potentials $V(r)$, Eq.~(\ref{eq:V}), for $\theta=0^\circ,20^\circ,40^\circ,60^\circ,80^\circ$, with the $\theta$-independent levels $\xit$ (dashed). (a) Kratzer-type, Eq.~(\ref{eq:VK}), with $qE=0.2$, $A=1$, $G=1/2$; levels for $l=0$, $k=0$--$3$. (b) Pseudoharmonic-type, Eq.~(\ref{eq:VP}), with $qE=1$, $\omega=1$, $G=1/2$, $c_0=0$; levels for $l=0$, $k=0$--$4$. (c) Hulth\'en-type with repulsive core, Eq.~(\ref{eq:VH}), with $qE=0.2$, $V_0=1/2$, $V_1=1/20$, $\delta=1/4$; the three $s$-wave levels. Atomic units.}
\label{fig:potentials}
\end{figure}

Figure~\ref{fig:potentials} shows the three field-modified central potentials for several orientations, together with the corresponding $\theta$-independent levels. In all three cases the repulsive core produces a minimum at finite $r$ whose position follows Eqs.~(\ref{eq:rminK}), (\ref{eq:rminP}) and (\ref{eq:rminH}); as $\theta$ decreases, the linear term steepens the outer wall and the minimum moves inward and upward. In the Hulth\'en-type case the well is narrow, with $r_{\min}\simeq0.73$ at $\theta=\pi/2$ from Eq.~(\ref{eq:rminH0}), while the outer wall is linear with a slope set by $qE\cos\theta$; the spectrum is finite and the highest level lies close to threshold. For every example the family of curves fans out with $\theta$ while the levels remain fixed, which is the hallmark of the construction~(\ref{eq:V}) and agrees with the TRA results of Ref.~\cite{Osunmusanmi2026}.

\begin{figure}[t]
\centering
\includegraphics[width=0.6\textwidth]{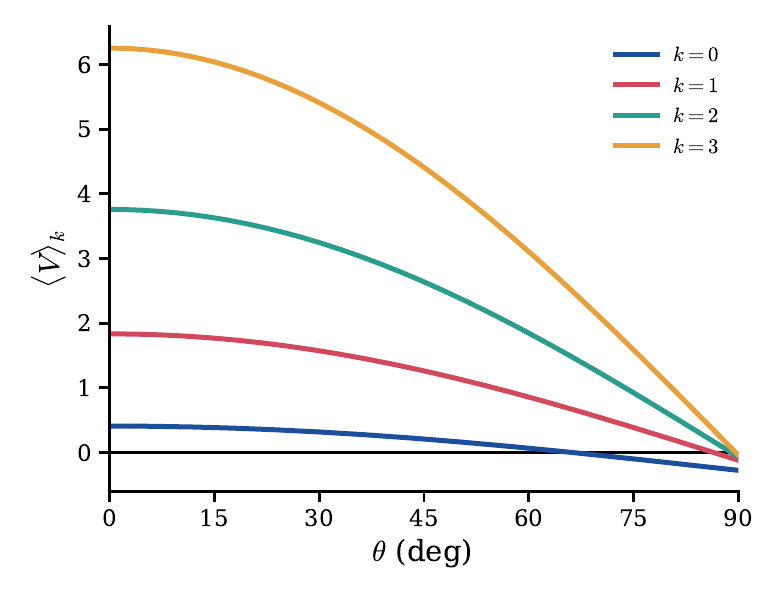}
\caption{Expectation value $\langle V\rangle_k$ of the Kratzer-type field-modified central potential, Eq.~(\ref{eq:VexpK}), versus $\theta$ for $l=0$ and $k=0$--$3$, with $qE=0.2$, $A=1$, $G=1/2$.}
\label{fig:Vexp}
\end{figure}

\begin{figure}[t]
\centering
\includegraphics[width=\textwidth]{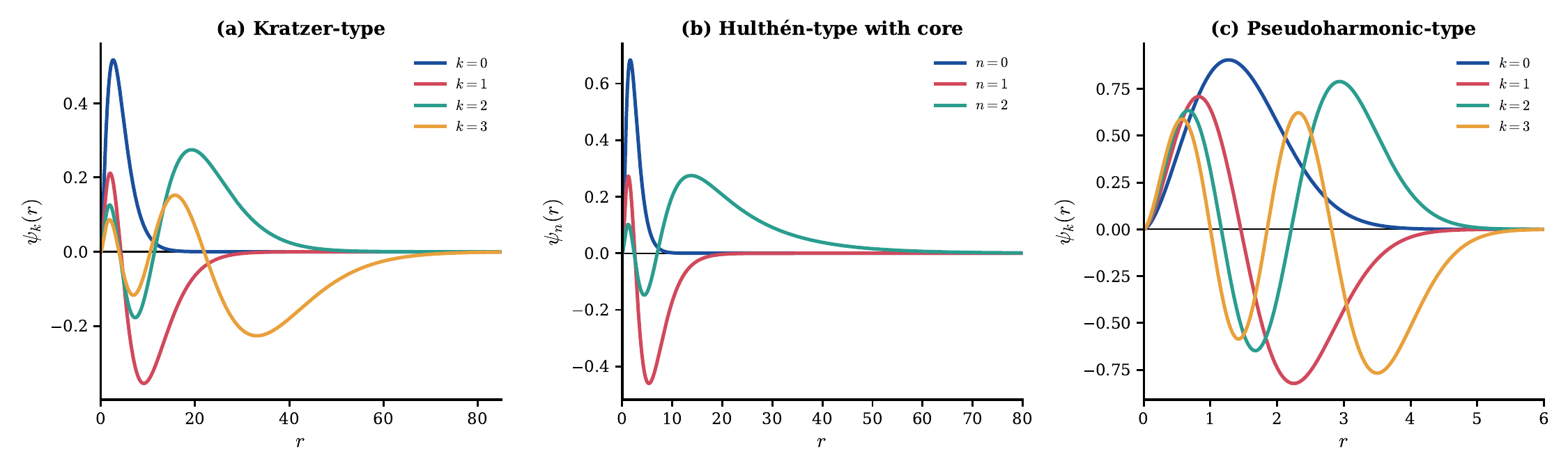}
\caption{Normalized eigenfunctions for $l=0$. (a) Kratzer-type, Eq.~(\ref{eq:psiK}), with $A=1$, $G=1/2$, $k=0$--$3$. (b) Hulth\'en-type with repulsive core, Eq.~(\ref{eq:psiH}), with $V_0=1/2$, $V_1=1/20$, $\delta=1/4$, for the three bound states $n=0,1,2$. (c) Pseudoharmonic-type, Eq.~(\ref{eq:psiP}), with $\omega=1$, $G=1/2$, $k=0$--$3$. The horizontal line marks $\psi=0$; the state with label $k$ (or $n$) has $k$ (or $n$) nodes. The eigenfunctions are independent of $\theta$.}
\label{fig:wf}
\end{figure}

Figure~\ref{fig:Vexp} displays $\langle V\rangle_k$ from Eq.~(\ref{eq:VexpK}) as a function of $\theta$. At $\theta=90^\circ$ the field term vanishes and $\langle V\rangle_k$ reduces to $\langle\Vt\rangle_k$, which is negative and approaches zero as $k$ grows. As $\theta$ decreases, the field contribution $qE\cos\theta\,\langle r\rangle_k$ grows as $(k+L+1)^2$, so the excited states are much more sensitive to the orientation than the ground state. This quantity, together with $r_{\min}(\theta)$, provides an analytical diagnostic of how the orientation modifies the potential sampled by each eigenstate, even though the eigenvalues themselves are unchanged. The normalized eigenfunctions~(\ref{eq:psiK}), (\ref{eq:psiH}) and (\ref{eq:psiP}) are shown in Fig.~\ref{fig:wf}. In the Kratzer-type and pseudoharmonic-type cases the inverse-square term replaces the integer $l$ by the noninteger $L$ in both the short-distance behavior $r^{L+1}$ and the Laguerre index, and the Kratzer-type outer lobes extend to $r\sim(k+L+1)^2/A$. In the Hulth\'en-type case the eigenfunctions vanish as $r^{\eta}$ at the origin; the highest bound state, $n=2$, lies close to threshold ($\epsilon_2\simeq0.35$) and is therefore much more extended than the deeply bound ground state.

Two further remarks are in order. First, the NU solutions do not rely on any tridiagonality requirement: once $\Vt(r)$ belongs to a class for which Eq.~(\ref{eq:radial}) is of hypergeometric type after a point transformation, the effective model is exactly solvable. The three examples therefore illustrate a general mechanism. Any other exactly solvable $\Vt(r)$, for instance of Mie or Eckart type within their respective exactness conditions, generates through Eq.~(\ref{eq:V}) a new orientation-dependent family of central potentials with a known spectrum. Second, the TRA and NU approaches are complementary. The TRA in its search mode constructs the admissible potentials from the basis and yields the expansion coefficients as orthogonal polynomials in the energy~\cite{Osunmusanmi2026,Alhaidari2017,Alhaidari2020}, whereas the NU method starts from a prescribed potential and yields the eigenfunctions as classical polynomials in the configuration variable. For the Coulomb-like and harmonic-like members, both lead to Laguerre-type structures, and the agreement of their structural predictions supports the consistency of the effective model.

%=====================================================================
\section{Conclusion}
\label{sec:conclusion}
%=====================================================================

We have shown that the effective field-modified radial Schr\"odinger equation introduced in Ref.~\cite{Osunmusanmi2026} can be solved exactly by a method other than the tridiagonal representation approach. The present work deliberately does not use the TRA; its aim is to demonstrate that the effective model is accessible to standard exact analytical methods. Using the Nikiforov--Uvarov method with prescribed Kratzer-type, pseudoharmonic-type and Hulth\'en-type components, we obtained three new families of orientation-dependent central potentials: a linear-plus-Coulomb-plus-inverse-square potential, a linear-plus-harmonic-plus-inverse-square potential, and a screened Hulth\'en--Manning--Rosen-type potential with a linear outer wall. For each family we derived the energy spectrum, the normalized eigenfunctions, the position of the potential minimum, and closed-form radial moments, including the orientation-dependent expectation value $\langle V\rangle_k$. All results were confirmed by independent numerical diagonalization.

In the present construction the eigenvalues are those of the radial operator containing $\acute{V}(r)$. The spectra are therefore independent of the orientation and of the electric field, in agreement with the TRA construction of Ref.~\cite{Osunmusanmi2026}; the electric field and the orientation enter through the effective central potential $V(r)=\acute{V}(r)+qEC\,r$, its minimum, and the radial expectation values, while the magnetic field enters the energies only through the linear Zeeman shift $qBm_\phi/2$.

Several extensions suggest themselves. Other exactly solvable components, such as Mie- and Eckart-type potentials, generate further families through the same mechanism, and nonzero angular momentum for screened potentials can be treated with controlled approximations. Supersymmetric and asymptotic-iteration techniques~\cite{Cooper1995,Ciftci2003} can be applied to the same effective model. Finally, if the linear term is retained in the radial operator rather than absorbed into $V(r)$, the spectrum acquires an explicit dependence on $qE\cos\theta$; for suitable choices of $\acute{V}(r)$ this leads to quasi-exactly solvable problems of biconfluent Heun type, which are treated in a companion paper~\cite{Osunmusanmi2026c}.

\section*{Acknowledgments}
This research was funded by Khalifa University of Science and Technology through Project ID KU-INT-RIG-2024-8474000739, with additional support from the Khalifa University Research Center for Cyber-Physical Systems (KU-C2PS). T.~J.~O. acknowledges the support of the Saudi Centre for Theoretical Physics, Dhahran, Saudi Arabia.

\subsection*{Conflict of interest}
The authors declare no potential conflict of interests.

\subsection*{Data availability statement}
The data that support the findings of this article are contained within the article.

\appendix
\setcounter{equation}{0}
\renewcommand{\theequation}{A\arabic{equation}}
\setcounter{table}{0}
\renewcommand{\thetable}{A\arabic{table}}

%=====================================================================
\section{Worked example: the Kratzer-type component}
\label{app:kratzer}
%=====================================================================

We give the intermediate steps of Sec.~\ref{sec:kratzer} as an illustration of how the NU method is applied. The pseudoharmonic-type and Hulth\'en-type cases follow the same pattern, with the data listed in Secs.~\ref{sec:pseudo} and~\ref{sec:hulthen}.

\paragraph{Perfect-square condition.}
For the data~(\ref{eq:NUK}) one has $(\sigma'-\tilde{\tau})/2=\tfrac12$. Using $\tfrac14+L(L+1)=\big(L+\tfrac12\big)^{2}$, the radicand in Eq.~(\ref{eq:pi}) becomes
\begin{equation}
R(r)=\epsilon^{2}r^{2}+(\kappa-2A)\,r+\left(L+\tfrac12\right)^{2}.
\label{eq:appR}
\end{equation}
A quadratic is a perfect square if and only if its discriminant vanishes, $(\kappa-2A)^{2}=4\epsilon^{2}\big(L+\tfrac12\big)^{2}$. This gives $\kappa_{\pm}=2A\pm\epsilon(2L+1)$, for which $R(r)=\big[\epsilon r\pm\big(L+\tfrac12\big)\big]^{2}$.

\paragraph{Selection of $\pi$.}
With the sign in front of the square root, there are four candidate functions $\pi$. The corresponding factors $\varphi=\exp\!\int(\pi/\sigma)\,dr$ are listed in Table~\ref{tab:candidates}. Only the last candidate yields a function that is regular at the origin and decays at infinity. Equivalently, it is the only choice for which $\tau=\tilde{\tau}+2\pi=2(L+1)-2\epsilon r$ has $\tau'<0$ and a zero, $r=(L+1)/\epsilon$, inside $(0,\infty)$, which is the NU selection rule. This gives Eq.~(\ref{eq:choiceK}).

\begin{table}[h]
\centering
\caption{The four candidate functions $\pi$ for the Kratzer-type component and the corresponding factors $\varphi=\exp\!\int(\pi/\sigma)\,dr$.}
\label{tab:candidates}
\begin{tabular}{cccc}
\hline
$\kappa$ & $\pi(r)$ & $\varphi(r)$ & Behavior \\
\hline
$\kappa_{+}$ & $\epsilon r+L+1$ & $r^{L+1}e^{+\epsilon r}$ & diverges at $\infty$ \\
$\kappa_{+}$ & $-\epsilon r-L$ & $r^{-L}e^{-\epsilon r}$ & singular at $0$ \\
$\kappa_{-}$ & $\epsilon r-L$ & $r^{-L}e^{+\epsilon r}$ & singular and divergent \\
$\kappa_{-}$ & $L+1-\epsilon r$ & $r^{L+1}e^{-\epsilon r}$ & accepted \\
\hline
\end{tabular}
\end{table}

\paragraph{Quantization.}
From $\lambda=\kappa+\pi'$ one finds $\lambda=2A-2\epsilon(L+1)$, while Eq.~(\ref{eq:lambdan}) with $\sigma''=0$ gives $\lambda_k=-k\tau'=2\epsilon k$. Equating the two expressions yields $\epsilon_k=A/(k+L+1)$, and $\xit=-\epsilon^{2}/2$ then gives Eq.~(\ref{eq:EK}).

\paragraph{Eigenfunctions.}
The factor $\varphi$ and the weight function follow from $\varphi'/\varphi=\pi/\sigma$ and $(\sigma\rho)'=\tau\rho$:
\begin{equation}
\varphi=r^{L+1}e^{-\epsilon r},\qquad
\frac{\rho'}{\rho}=\frac{2L+1}{r}-2\epsilon\;\Rightarrow\;\rho=r^{2L+1}e^{-2\epsilon r}.
\end{equation}
The Rodrigues formula~(\ref{eq:rodrigues}) then gives
\begin{equation}
y_k=B_k\,r^{-(2L+1)}e^{2\epsilon r}\frac{d^{k}}{dr^{k}}\Big[r^{k+2L+1}e^{-2\epsilon r}\Big].
\end{equation}
With $x=2\epsilon_k r$, this coincides, up to a constant, with the Rodrigues formula of the associated Laguerre polynomials,
\begin{equation}
L_k^{(\alpha)}(x)=\frac{x^{-\alpha}e^{x}}{k!}\frac{d^{k}}{dx^{k}}\Big(x^{k+\alpha}e^{-x}\Big),\qquad \alpha=2L+1,
\end{equation}
so that $\psi=\varphi\,y_k$ is given by Eq.~(\ref{eq:psiK}).

\paragraph{Normalization.}
With $x=2\epsilon_k r$ and $\alpha=2L+1$,
\begin{equation}
\int_0^\infty\psi_{k,l}^{2}\,dr=\frac{N_k^{2}}{(2\epsilon_k)^{2L+3}}\int_0^\infty x^{\alpha+1}e^{-x}\Big[L_k^{(\alpha)}(x)\Big]^{2}dx .
\label{eq:appnorm}
\end{equation}
Because the power $x^{\alpha+1}$ differs from the Laguerre weight $x^{\alpha}e^{-x}$, the orthogonality relation
\begin{equation}
\int_0^\infty x^{\alpha}e^{-x}L_n^{(\alpha)}L_m^{(\alpha)}\,dx=\frac{\Gamma(n+\alpha+1)}{n!}\,\delta_{nm}
\label{eq:apporth}
\end{equation}
cannot be used directly. Instead, we multiply the recurrence relation
\begin{equation}
x\,L_n^{(\alpha)}=(2n+\alpha+1)L_n^{(\alpha)}-(n+1)L_{n+1}^{(\alpha)}-(n+\alpha)L_{n-1}^{(\alpha)}
\end{equation}
by $x^{\alpha}e^{-x}L_n^{(\alpha)}$ and integrate. By Eq.~(\ref{eq:apporth}) the terms with $L^{(\alpha)}_{n\pm1}$ vanish, and
\begin{equation}
\int_0^\infty x^{\alpha+1}e^{-x}\Big[L_n^{(\alpha)}\Big]^{2}dx=(2n+\alpha+1)\,\frac{\Gamma(n+\alpha+1)}{n!}.
\end{equation}
For $n=k$ and $\alpha=2L+1$ the prefactor is $2(k+L+1)$, and setting Eq.~(\ref{eq:appnorm}) equal to unity gives Eq.~(\ref{eq:NK}). For $G=0$ one has $L=l$ and $\epsilon_k=A/n$ with $n=k+l+1$, and Eq.~(\ref{eq:NK}) reduces to the standard hydrogen-like normalization $N^{2}=(2A/n)^{2l+3}(n-l-1)!/[2n\,(n+l)!]$.

\paragraph{Jacobi case.}
For $\sigma=z(1-z)$, as in Sec.~\ref{sec:hulthen}, the Rodrigues formula generates Jacobi polynomials in the variable $1-2z$~\cite{Nikiforov1988,Koekoek2010}. Comparing Eq.~(\ref{eq:hyp}) with the Jacobi equation identifies $\tau=a+1-(a+b+2)z$, which gives the indices $a=2\epsilon_n$ and $b=2\eta-1$ and the eigenvalue $n(n+a+b+1)=n(n+2\epsilon_n+2\eta)$, in agreement with $\lambda_n$.

\end{document}